# Complex Octahedral Tilt Phases in the Ferroelectric Perovskite System $Li_xNa_{1-x}NbO_3$

Charlotte A. L. Dixon and Philip Lightfoot*

School of Chemistry and EaStCHEM, University of St Andrews, St Andrews, KY16 9ST, UK

*E-mail **pl@st-and.ac.uk**

**Abstract**

High-temperature phase behavior in the system $Li_xNa_{1-x}NbO_3$ has been studied by using high-resolution neutron powder diffraction. Each of the three compositions studied in the Na-rich part of the phase diagram (*viz.* x = 0.03, 0.08 and 0.12) shows evidence for distinct and complex structural modulations based on different tilting schemes of $NbO_6$ octahedral units. Whilst octahedral titling is prevalent in the structural chemistry of perovskites the nature and complexity of the phases observed here is unprecedented. Neither of the long-range tilt phases previously observed in $NaNbO_3$ itself occurs here; instead a novel phase with a well-defined 4-fold superlattice is observed for the composition $Li_{0.12}Na_{0.88}NbO_3$, and yet more complex phases with modulations based on 20-fold and 30-fold repeats are observed for $Li_{0.03}Na_{0.97}NbO_3$ and $Li_{0.08}Na_{0.92}NbO_3$, respectively. This peculiar structural frustration makes the system $Li_xNa_{1-x}NbO_3$ the most structurally complex 'simple' perovskite known.

## Introduction

The search for a less toxic, lead-free alternative to the piezoelectric material PZT ($PbZr_{1-x}Ti_xO_3$) has received much attention in recent years, with alkali niobate perovskites offering some promising candidate materials[1]. Of these, materials based on the KNN ($K_xNa_{1-x}NbO_3$) solid-solution have proven most promising[2]. The composition-dependent and temperature-dependent phase diagrams of KNN have consequently been widely studied[3,4]. The corresponding phase behavior and crystallography of the LNN system ($Li_xNa_{1-x}NbO_3$) have been much less well explored, despite both end-members having structural and/or physical properties of significant interest and uniqueness: $LiNbO_3$ displays excellent electro-optical properties[5], whereas $NaNbO_3$ has one of the most complex temperature-dependent phase diagrams of a simple $ABO_3$ perovskite[6,7]. $LiNbO_3$, on the other hand, show much simpler temperature-dependent structural behavior[8]. A detailed study of the phase behavior of LNN as a function of both composition and temperature therefore seems of fundamental importance, especially in the light of conflicting reports of phase behavior and physical properties, in particular of enhanced piezoelectric properties around the composition $Li_{0.12}Na_{0.88}NbO_3$[9,10]. We have previously carried out a thorough study of the phase behavior of LNN as a function of composition at room-temperature, and concluded that synthesis conditions, such as annealing temperature and cooling rate, could significantly affect the phases present[11]. Moreover, samples at certain compositions are even susceptible to phase transformations simply standing in air at ambient conditions for extended periods. The overall conclusion of that study is that sample preparation conditions and sample characterization must be carefully carried out and reported, if reproducible physical properties are to be derived. A schematic phase diagram for LNN at ambient temperature is shown in Figure 1[11].

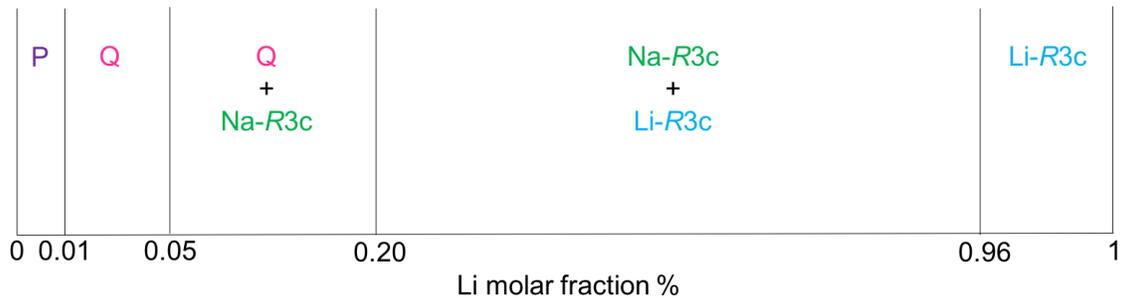

**Figure 1:** Ambient temperature phase diagram for the $Li_xNa_{1-x}NbO_3$ system. Phase abbreviations are defined in the text, and in ref. 11.

We have now extended the crystallographic characterization of LNN to the variable temperature regime. For the composition $Li_{0.20}Na_{0.80}NbO_3$[12] we recently reported a highly unusual phase transition sequence at elevated temperature, involving five consecutive phases, two of which are based on the very rare Glazer octahedral tilt system[13] $a^+a^+c^-$. In the present work, we find that even more complex behavior is observed towards the higher Li content region of the phase diagram (x < 0.20). In this study we analyze in detail the thermal evolution of three selected compositions within this Na-rich part of the phase diagram. We use high resolution powder neutron diffraction (PND) to study the compositions $Li_xNa_{1-x}NbO_3$ for x = 0.03, 0.08 and 0.12 (hereafter abbreviated LNN-3, LNN-8, LNN-12, respectively). A recent paper[14] has carried out a similar study on a single composition, at LNN-12, and our results will be compared.

**Experimental Section**

**Synthesis:** LNN ceramics were synthesized using a conventional mixed oxide and carbonate solid-state route. Samples of $Li_xNa_{1-x}NbO_3$ were prepared with x = 0.03, 0.08 and 0.12. Stoichiometric amounts of $Na_2CO_3$ (99.9%, Sigma), $Li_2CO_3$ (99.9%, Sigma) and $Nb_2O_5$ (99.9%, Alfa Aesar) were dried at 100 °C for 48 hours, ground for 30 minutes, and pressed into pellets of 10 mm diameter and approximately 2 mm thickness. Samples were subsequently heated at temperatures between 950 °C for LNN-3 and LNN-8 and 1100 °C for LNN-12 for 24 hours, and cooled at a rate of 10 °min$^{-1}$. Pellets were then re-ground to produce final powders suitable for characterization. All X-ray diffraction data are obtained within 1 day of synthesizing a sample, and neutron diffraction data within 10 days.

**Powder X-ray diffraction (PXRD):** Powdered samples were loaded into disks of 3 mm depth, and data were obtained using Cu K$_{\alpha 1}$ radiation with a wavelength of 1.5406 Å. Diffraction patterns were obtained using a Panalytical EMPYREAN diffractometer with a 2θ step size of 0.017°. All data were recorded at room temperature with a 1 hour scan time.

**Powder neutron diffraction (PND):** Time-of-flight neutron powder diffraction experiments were conducted using the high-resolution powder diffractometer (HRPD) at the ISIS neutron spallation source at the Rutherford-Appleton Laboratories. The polycrystalline samples (~3 g) were mounted in cylindrical vanadium cans. Data were collected at a series of temperatures commencing at 20 °C, and subsequently at 50 ° intervals for the temperature regime $50 \leq T \leq 900$ °C. Each scan was counted for 30 µAhr incident proton beam (ca. 60 min).

**Diffraction Data Analysis:** All diffraction data were analyzed by Rietveld refinement using the GSAS software package with the EXPGUI interface. Refinement strategies were kept as uniform as possible across all datasets. For each dataset, two diffraction histograms were used (detector banks centered at $2\theta = 168°$ (bank 1) and $90°$ (bank 2)). The same set of profile parameters were refined in each case: nine background coefficients and three peak shape parameters for each histogram, three diffractometer constants and two histogram scale factors in total. In addition to these common parameters, appropriate lattice parameters, all atomic positional coordinates and all isotropic atomic displacement parameters (ADPs, grouped by element type). For each composition the Na/Li ratio was fixed and these atoms were treated with equivalent atomic coordinates and ADPs. Further details of refinement strategies are given in the relevant sections, as necessary. Symmetry mode analysis was carried out using the ISODISTORT suite[15].

**Results**

We have analyzed in detail the PND data on three distinct sample compositions, LNN-3, LNN-8 and LNN-12, at all temperatures recorded, covering a total of 57 separate datasets. Due to the many distortion modes possible in perovskites[7,11,16,17], many of the datasets require consideration of several distinct, but often very similar, crystallographic models. In order to allow a clear and systematic presentation of these results, and their interpretation, we have chosen to commence by describing the behavior of LNN-12, for two reasons. First, this composition has been the one most widely studied in previous literature and, second, it nicely illustrates some of the complexities of data analysis and, ultimately, reveals clear and unambiguous evidence of a novel perovskite structural variant. The following discussion highlights the key phases identified, and their evolution versus temperature. Full details of each of the refinement models considered, together with further supporting Rietveld plots, are provided in the Supplementary Information (SI).

**LNN-12:**

At ambient temperature, lab PXRD suggests that the as-prepared LNN-12 corresponds to the polar rhombohedral phase (designated 'phase N' in the early literature, space group $R3c$), which we abbreviate as 'Na-R3c'[11]. Our PND study commences at 50 °C. On close inspection of the raw data it can be seen that a trace amount of the polar orthorhombic 'phase Q'[6,18] (space group $P2_1ma$) is also present; this was observed in our previous work, where phase co-existence was shown to be subtlely dependent on preparation conditions[11]. Rietveld refinement (Fig. 2(a)) reveals 91 % Na-R3c and 9 % phase Q at this temperature. On raising the temperature to 100 °C, it can be seen that the phase fraction of Q increases (47%), at the expense of Na-R3c (53%), and subsequently at 150 °C only phase Q is present (Fig. 2(b)). In Glazer notation[13], phase Na-R3c has an octahedral tilt system $a^-a^-a^-$, whereas phase Q has the tilt system $a^-b^+a^-$. These are the most common octahedral tilt systems, but in these particular cases the tilt systems are augmented by an additional polar mode (crystal structures for both phases are given in the SI, Fig. S1). In order to confirm that the phase observed here is indeed the polar phase Q, rather than the ideal centrosymmetric $a^-b^+a^-$ phase, a comparative refinement was carried out for the 150 °C dataset in the *Pnma* space group, which is the aristotype symmetry for this tilt system. This resulted in a considerably poorer quality of fit ($P2_1ma$: $\chi^2 = 2.18$ for 56 variables; *Pnma*: $\chi^2 = 3.71$ for 43 variables). On raising the temperature, the continued presence of pure Q is confirmed up to 250 °C; at 300 °C detailed examination of the raw PND data plots shows the presence of a small amount of a new phase, which continues to grow-in towards 350 °C, and becomes a single phase at 400 °C (Fig. 3). A detailed characterization of the phase present at 400 °C is now described.

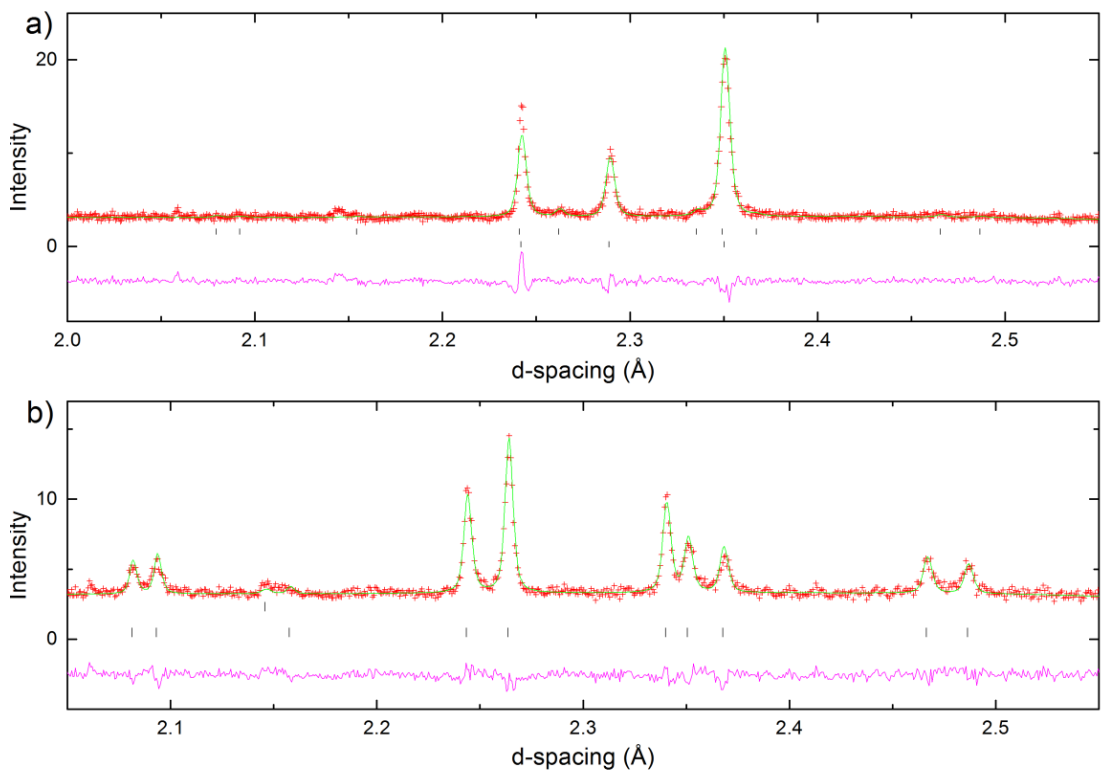

**Figure 2:** Rietveld refinement of PND data for LNN-12 taken from bank 1, (a) measured at 50 °C and modelled as a phase co-existence of Q and Na-R3c phases. The top set of Bragg peaks correspond to Phase Q and the bottom set to Na-R3c. $\chi^2 = 2.11$, $R_{wp} = 0.0383$; and (b) measured at 150 °C and modelled as phase Q only. $\chi^2 = 2.18$, $R_{wp} = 0.0391$. The peak at d ~ 2.15 Å originates from the vanadium canister.

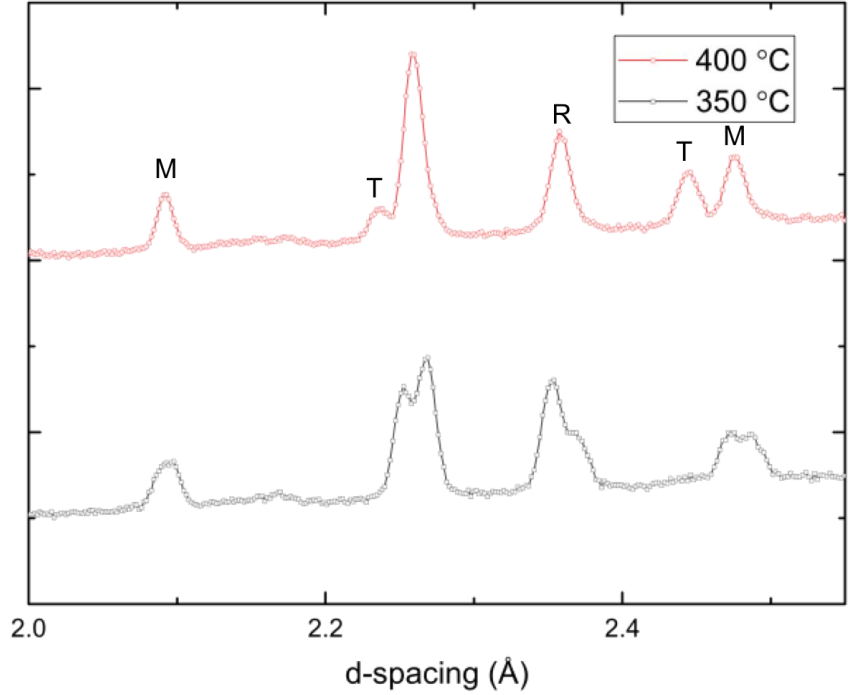

**Figure 3:** Raw PND data plots (bank 2) measured at 350 °C (black) and 400 °C (red) for LNN-12, highlighting the completed transition to the new phase at 400 °C, exhibiting *T*-point peaks.

Fig. 3 highlights several key peaks which correspond to scattering at specific points in the parent (pseudo)-cubic 1st Brillouin zone. The *M*-point, $k$ = (½, 0, ½) and *R*-point, $k$ = (½, ½, ½) peaks arise, at least in part, from in-phase ('+'), and out-of-phase ('-') octahedral tilts; corresponding peaks are seen in phase Q at lower temperatures (Fig. 2(b)), in which case these tilts act along the [1,0,0] / [0,0,1] and [0,1,0] directions of the parent cubic unit cell, respectively. The two new peaks, labelled '*T*', arise from a distinctly different distortion, which acts at a specific point $k$ = (½, ½, ¼) along the *T*-line, $k$ = (½, ½, γ) of the cubic Brillouin zone. Our assumption in this analysis is that these peaks also, *at least in part*, derive from octahedral tilt modes. In our previous work[7] on the high temperature phases of NaNbO$_3$ itself, we carried out a much more exhaustive analysis of related phases, without making this assumption, but subsequently demonstrated that the corresponding peaks are indeed largely tilt-induced. As will be shown, this assumption provides a useful starting point for the present analysis. The specific tilt modes which contribute significantly to the peaks observed at the *M*, *R* and *T*-points are designated by the *irreps* $M_3^+$, $R_4^+$ and $T_4$.

The presence of peaks at the *T*-point $k$ = (½, ½, ¼) arises from a quadrupling of the aristotype cubic perovskite unit cell parameter, $a_p$ (where $a_p$ ~ 3.9 Å). Combined with the presence here of the *M* and *R* peaks, this requires a minimum unit cell metric for the present phase of ~ √2 $a_p$ ×√2 $a_p$ × 4 $a_p$. This is reminiscent of Phase S of pure NaNbO$_3$, which we characterized in our previous work[7]. In that work we carried out a systematic and exhaustive search for all distortions of the parent cubic perovskite, which give rise to this type of cell metric, and also the larger 2 $a_p$ × 2 $a_p$ × 4 $a_p$ supercell option. By filtering and testing a bewildering array of options, we were able to derive several models which fitted the observed PND patterns very closely, but only one of which was deemed a fully satisfactory model. This highlights a key challenge in deciphering the details of complex distortions in perovskites; that of pseudo-symmetry. The presence of the *M*, *R* and *T*-point peaks reveal only the presence of particular *types* of tilt mode, but not their specific *direction* relative to the unit cell axes. In the case of severe overlap of unit cell metrics, which can be 'coincidental' rather than symmetry-dictated, it is very challenging to assign the associated directions and, without the availability of the highest possible resolution, almost beyond the capabilities of powder diffraction methods. Here, we show that, with careful analysis, based on systematic consideration of the available options, it is possible to derive a unique model amongst many very similar ones.

We used the on-line software ISODISTORT[15] to generate possible starting models, based on the simultaneous presence of the three distinct types of octahedral tilt, $M_3^+$, $R_4^+$ and $T_4$. We note that there are two distinct $T_4$ tilt normal modes (or *irreps*): we use the notation from ref. 7 to distinguish these as AACC and A0C0 (C = clockwise tilt, A = anti-clockwise tilt of equivalent absolute magnitude). These are shown in Fig. S2. Coupling of these three types of *irrep* (*i.e.* $M_3^+$, $R_4^+$ and $T_4$) produces a total of 96 distinct starting models (see SI, for full details). Of these, many are most likely to be of unnecessarily low symmetry. In passing, it is interesting to note that both non-centrosymmetric and even polar subgroups arise naturally from the combination of these three tilt modes, whereas distorted perovskites produced from the combination of $M_3^+$ and $R_4^+$ tilts only are necessarily centrosymmetric[17]. Combinations of distortion modes, for example based on tilts and cation ordering, have previously used as a 'design principle' for induction of ferroelectricity in perovskites[19,20], so the possibilities of combining complex tilts in this way may also be of future interest. In order to produce a manageable set of models, which nevertheless, incorporate all of the most reasonable simple combinations of tilt modes, we limit our consideration to models which are (a) centrosymmetric and (b) of index $i = 96$ or lower (the sub-group index $i$ is effectively a measure of the 'dilution' of the symmetry operations of the subgroup compared to the parent group; i.e. $i = (M_p/M_s) \times (V_s/V_p)$, where $M_p$ and $M_s$ are the multiplicities of a general site in the parent and subgroup unit cells, and $V_p$ and $V_s$ the corresponding unit cell volumes). These nine models are shown in Table I. Of these, we chose a further subset of six models for detailed testing, the idea being to include examples of all simple ways to superimpose crystallographic directions for three distinct tilt *irreps*. For example, the model LNN-S1 exhibits the $M_3^+$, $R_4^+$ and $T_4$ tilts simultaneously along the *c*-axis, whereas models LNN-S3 and LNN-S8 require each of the three tilts to occur relative to three different axes but differ in the nature of the $T_4$ irrep. Note that Table I incorporates each of the key models we considered (and derived through an exhaustive search, from quite different assumptions) in ref. 7; the monoclinic model LNN-S4 was not considered in our previous work.

Table 1[$]. Trial models for LNN-12 at 400 °C ("Phase S"-like) derived from superposition of $M_3^+$, $R_4^+$ and $T_4$ tilts (see text).

| Model | Space group | $N_{xyz}$ | Metrics | Origin | $R_4^+$ $M_3^+$ $T_4$ | $T_4$ | $\chi^2$ |
|---|---|---|---|---|---|---|---|
| LNN-S1 | *P4/mbm* | 9 | √2 √2 4 | (0,0,0) | *c c c* | A0C0 | 11.2 |
| LNN-S2 | *Pbcm* | 29 | 2 4 2 | (0,0,0) | *a a b* | A0C0 | 4.74 |
| **LNN-S3** | ***Pnma*** | **31** | **2 2 4** | **(0,0,0)** | ***a b c*** | **A0C0** | **2.61** |
| LNN-S4 | *C2/m* | 30 | 2 2 4 | (1/2,1/2,0) | *(a,c) c c* | A0C0 | ~4 |
| LNN-S5 | *Pmmn* | 33 | 2 2 4 | (0,0,0) | *c (a,b,c) c* | A0C0 | 3.04 |
| LNN-S6 | *Pmma* | | 4 2 2 | (-1/2,1/2,0) | *c c a* | AACC | |
| LNN-S7 | *Pmma* | | 2 2 4 | (-1/2,0,1/2) | *c a c* | AACC | |
| **LNN-S8** | ***Pmmn*** | **35** | **4 2 2** | **(1/2,0,1/2)** | ***c b a*** | **AACC** | **2.45** |

| LNN-S9 | $P2_1/c$ | 4 | $\sqrt{2}$ | $\sqrt{2}$ | (0,0,0) | *(bc, a) c c* | A0C0 |

$N_{xyz}$ is the number of allowed variable atomic coordinates; in addition, three ADPs, relevant lattice parameters and the same set of 29 profile parameters were refined in each case. 'Metrics' and 'Origin' define the settings of the supercell metrics and the origin choice relative to the parent cubic phase (space group $Pm\bar{3}m$: Nb at (0,0,0); Na at (1/2,1/2,1/2); O at (1/2,0,0)). The sixth column shows the supercell crystallographic axes along which the three principal tilt modes act; the seventh column states which of the two possible $T_4$ irreps are active. The refinement for the LNN-S4 did not converge.

First of all, the model found to be the most satisfactory for Phase S in our previous study of $NaNbO_3$ was tested here (model S-P12a in ref. 7 and model LNN-S5 in Table I). Using a careful damping strategy, it was found possible to achieve a stable Rietveld refinement incorporating free refinement of all atomic coordinates, lattice parameters, peak shape parameters and isotropic ADPs, grouped by element type. Although a very good quality fit was achieved, closer inspection revealed a slight misplacement of the fit to the *T*-point peak near $d = 2.44$ Å (Fig. 4(a)). This suggests that this model, which requires both the $R_4^+$ and $T_4$ tilts to occur along the same axis, is probably incorrect for this composition. Subsequently, each of the other five models highlighted in Table 1 were tested against the experimental data at 400 °C, by using ISODISTORT to derive starting coordinates based on approximate amplitudes (Å) of the three tilt modes: $M_3^+ = 1.4$, $R_4^+ = 1.4$ and $T_4 = 1.2$. These values were chosen as they represent the approximate observed amplitudes found for the final refinement of the LNN-S5 model described above; they were found to give sufficiently good starting approximations to the observed diffraction intensities at the *M*, *R* and *T*-points for each of the other cases. For each of these five models, the same type of unconstrained refinement of the relevant set of atomic parameters and profile parameters was also attempted. Final details of these refinements are given in Table I. As can be seen, two models stands out as significantly better than the rest; models LNN-S3 and LNN-S8. Both models require each of the three tilts to occur along *different* crystallographic axes (these correspond to models S-P17 and S-P13 in ref. 7). The key difference between the models is the nature of the $T_4$ tilt mode (AACC versus A0C0); it is impossible to suggest one of these models is significantly better than the other, given the number of variables versus quality of fit (Fig. 4). In any case, this analysis suggests, quite unambiguously, that the phase present here, although displaying some very similar characteristics to phase S of NaNbO$_3$ (specifically the presence of the same three tilt *irreps*), is actually distinct and unique in its overall tilt system. We designate this new phase, Phase S´, the crystal structure of which, for the LNN-S3 model, is shown in Fig. 4(b) (see also Table S1). The refined unit cell parameters for model LNN-S3 at 400 °C are $a$ = 7.82835(15), $b$ = 7.8196(2), $c$ = 15.6409(4) Å. Despite the similarity (within ~0.1%) of the reduced unit cell parameters (i.e. $a$/2, $b$/2, $c$/4), this difference is apparently sufficient to distinguish between the models with differing tilt orientations from the high quality powder diffraction data used.

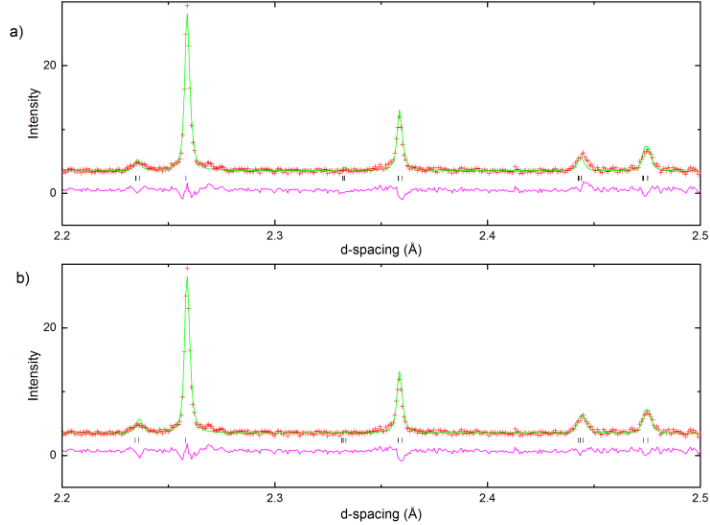

**Figure 4:** Portion of the Rietveld refinement obtained for LNN-12 at 400 °C, modelled as a) Phase S and b) Phase S'. Note the poor fit of the *T*-point peak near $d = 2.44$ Å in the Phase S model.

Phase S' remains present from 400 °C to 600°C; a much wider stability range than that observed for Phase S of NaNbO$_3$ itself (~ 480 °C < $T$ < 510 °C). In NaNbO$_3$ another complex phase, R, with a 2 $a_p$ × 2 $a_p$ × 6 $a_p$ metric, occurs in the lower temperature interval ~ 370 °C < $T$ < 470 °C. There is no evidence for such a phase in LNN-12 from the present data. In NaNbO$_3$ the phase sequence above the regime of Phase S is Phase T1 → Phase T2 → Phase U. These phases have 'standard' distortions[13,17] of the perovskite unit cell, with Glazer tilt notations and space groups, respectively: T1 ($a^0b^+c^-$; *Cmcm*); T2 ($a^0a^0c^+$; *P4/mbm*); U ($a^0a^0a^0$; $Pm\bar{3}m$). We find the same sequence of phases here for LNN-12, with orthorhombic phase T1 existing at 650 °C, tetragonal phase T2 at 700 °C and cubic phase U at 750 °C to 900 °C. These transitions can be seen clearly from the raw data plots in Fig. 5, where sequential disappearance of the $T_4$, $R_4^+$ and $M_3^+$ tilt-related peaks can be seen. The T1 → T2 and T2 → U transitions are allowed to be 2$^{nd}$ order according to Landau theory[17], and no evidence for phase co-existence is found in our refinements, in agreement with theory. The only other consideration is to check the alternatives to the T1 phase model. Two further standard models allow simultaneous presence of M and R-point tilts: $a^+a^+c^-$ (space group *P4$_2$/nmc*) and $a^-b^+a^-$ (space group *Pnma*). Fits were carried out against these models, for comparison to the *Cmcm* model, which show convincingly that *Cmcm* is indeed the correct model at 650 °C ($\chi^2$ values for *Cmcm*, *P4$_2$/nmc* and *Pnma* models respectively 1.68 (45 variables), 3.09 (44 variables) and 3.57 (43 variables)). Refined models and representative Rietveld plots for these higher temperature phases are provided in Fig. S3-5 and Table S2-4

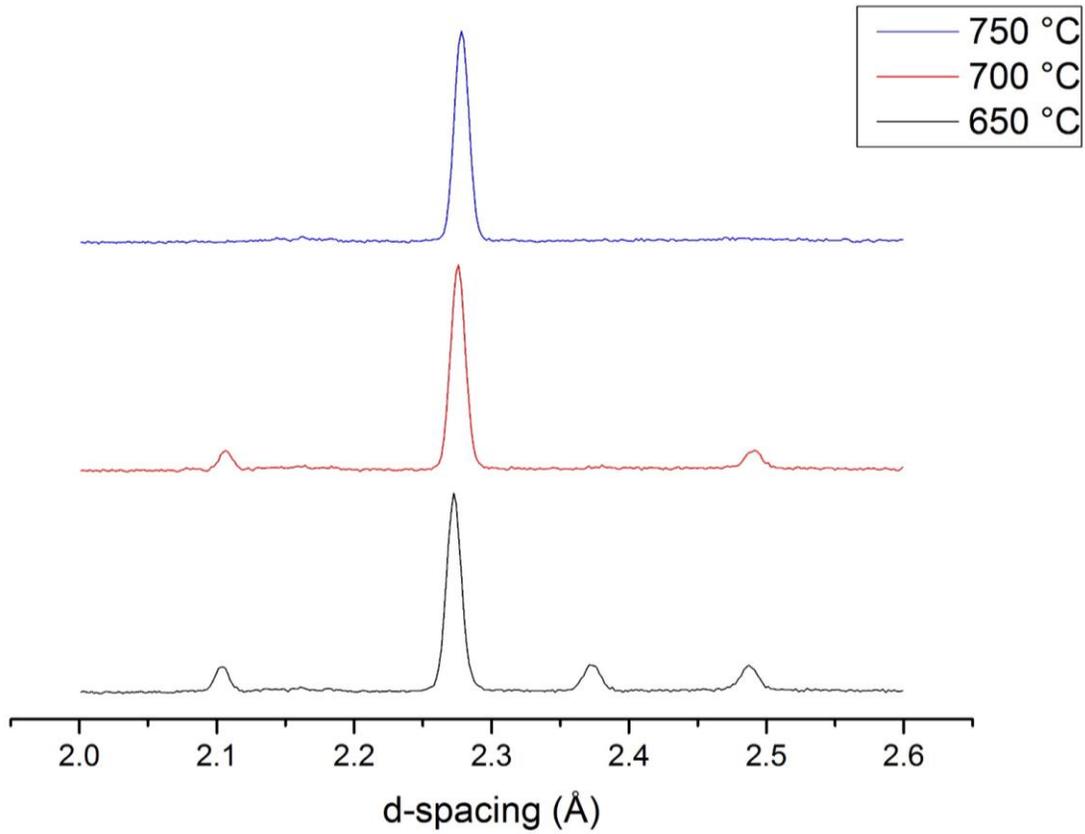

**Figure 5:** Raw data plots (bank 2) for LNN-12 showing the T1, T2 and U phases at 650 (black), 700 (red) and 750 °C (blue) respectively. Sequential loss of the $T_4$, $R_4^+$ and $M_3^+$ tilt-related peaks can be seen (compare Fig. 3).

**LNN-8:**

At temperatures from ambient up to 100 °C LNN-8 occurs as a mixture of Phase Q plus a small amount of the Na-R3c phase (about 4% Na-R3c at 50 °C). At 150 °C the small amount of Na-R3c has disappeared, and a good fit to the Phase Q model is obtained at temperatures up to 300 °C (Fig. S6). In the temperature regime 350 °C < $T$ < 600 °C the subcell lattice metrics converge and a new superlattice peak near $d \sim 2.45$ Å appears, at first sight corresponding to the *T*-point $\mathbf{k} = $ (½, ½, ¼). Co-existence of Q plus this S–like phase occurs at 350 °C, but the S-like phase is isolated at 400 °C. However, attempts to fit this new peak using either the Phase S or Phase S´ models proved unsuccessful (Fig. 6). These two models allow the options of *R*, *M* and *T* tilts along *different* axes (S´) and *R* and *T* tilts along the *same* axis (S). Obviously, the specific cell metrics for each of these models differ (for S, $a = 7.8287(2)$, $b = 7.8388(2)$, $c = 15.6659(6)$ for S´, $a = 7.8394(2)$, $b = 7.8300(3)$, $c = 15.6642(6)$) but regardless of which of these models is used, the underlying perovskite subcell metrics ($a_p$) are essentially unchanged. The three values ($a_p \sim 3.915$, 3.916, 3.920 Å) are very similar to each other, which emphasizes the challenges of assigning particular directions to each tilt. By using the S´ model but additionally permuting all possible directions of these 'long', 'short' and 'medium' subcell axes, it can be shown that none of these options give a satisfactory fit to the position of the *T*-point peak near $d \sim 2.45$ Å. Hence we conclude that the superlattice along at least one of the subcell directions is not a 4-fold repeat, and other points along the *T*-line need to be considered.

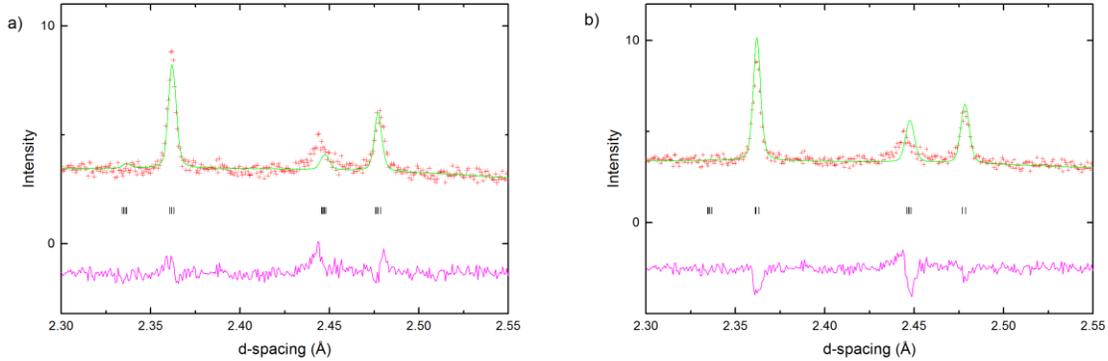

**Figure 6**: Rietveld refinement for LNN-8 recorded at 400 °C, showing unsatisfactory fitting of the *T*-point peak (near $d \sim 2.45$ Å) using a Phase S model (a) and Phase S' model (b).

We chose a systematic method of testing for various possible higher-order superlattices along the *T*-line. Using ISODISTORT, simple models of superlattice size $\sqrt{2}\, a_p \times \sqrt{2}\, a_p \times n\, a_p$ were derived, which contained the minimum requirement of a $T_4$ tilt mode. For $n = 3$ to $n = 10$ it turns out that the simplest possible models all correspond to tetragonal space

*P4/mbm*, with the same origin choice as the parent cubic perovskite. In each case, powder neutron diffraction patterns were simulated by switching on the relevant $T_4$ modes only, and the positions of the resulting *T*-point peaks were noted. The Miller indices of the peaks associated with tilt modes are of the form (3/2,1/2,$\gamma$) relative to the cubic aristotype subcell, or (3,1,*l*) relative to a 2 $a_p$ × 2 $a_p$ × n $a_p$ supercell; these peaks run along the *T*-line in reciprocal space, joining the *R*-point ($\gamma = ½$, $l = n/2$) to the *M*-point ($\gamma = 0$, $l = 0$). The *T*-points at $k = (½, ½, ¼)$ and $k = (½, ½, 2/7)$ give rise to tilt peaks close to the observed peak. However, they do not quite fit, the former occurring just to the higher *d*-spacing side, and the latter to the lower *d*-spacing side. The natural outcome of this is to propose a model with a longer range superlattice. A *T*-point at $k = (½, ½, 4/15)$ provides an excellent fit (Fig. 7 shows the d-spacing intensity of the $T_4$ mode with $\gamma = 2/7$, 4/15 and ¼).

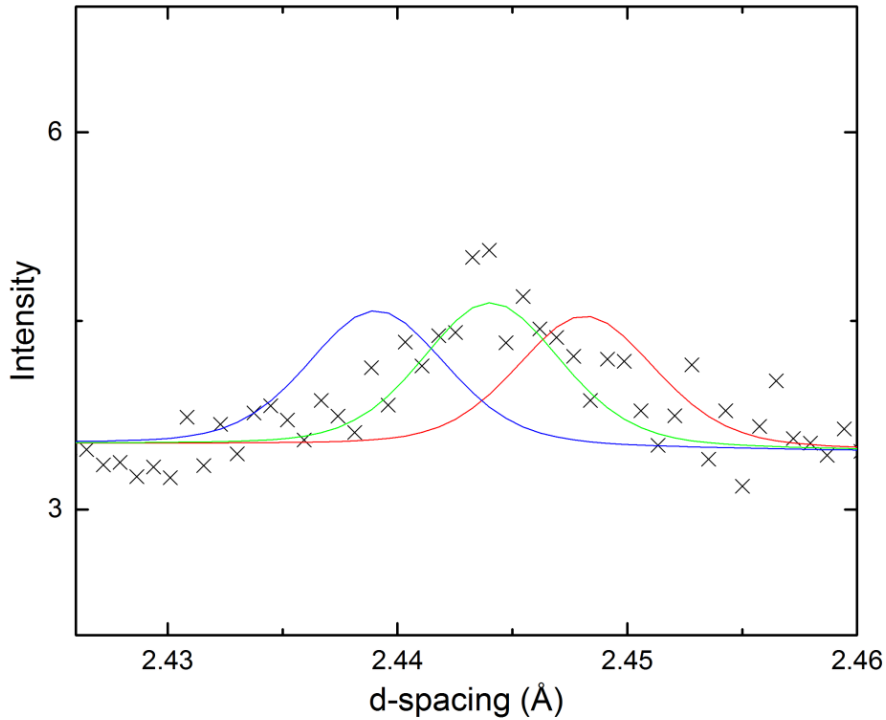

**Figure 7:** Histograms generated using the ISODISTORT software and a *P4/mbm* model with $T_4$ modes occurring at $k = (½, ½, \gamma)$, where $\gamma = 2/7$ (blue), 4/15 (green), 1/4 (red). The observed peak is depicted with black crosses.

Although the simplest model that fits the position of this peak correctly therefore requires a unit cell size of approximately $\sqrt{2}\, a_p \times \sqrt{2}\, a_p \times 15\, a_p$, and a space group $P4/mbm$, the unit cell actually needs to be bigger and of lower symmetry in order to simultaneously accommodate both the $M_3^+$ and $R_4^+$ tilt modes. In particular, the $R_4^+$ mode can only occur in unit cells with even numbered repeats of $a_p$. The simplest model that can account for the observed $R$, $M$ and $T$-point peaks hence requires a minimum cell size approximately $2\, a_p \times 2\, a_p \times 30\, a_p$. We designate this new proposed phase, Phase S´´. Of course, the number of options within such a large supercell precludes any meaningful attempt to propose a full and unique structural model from the present data. A plausible, contributory $T_4$ tilt *irrep* is shown in Fig. S7 but we note that significantly different tilts along the long axis can be constructed by combining different $T_4$ modes, together with the $R_4^+$ and $M_3^+$ modes; these have differing effects on overall peak intensities, but further analysis has not been pursued. On raising the temperature to 600 °C the $T$-point peak disappears, but the $M$- and $R$-point peaks remain. Phase T1 (*Cmcm*) provides a good fit. At 650 °C the $M$-point peak is still strong, but the $R$-point has diminished to near-zero, suggesting a likely transition to phase T2 (*P4/mbm*). Unfortunately, this intermediate phase cannot be confirmed, as the next available data point (700 °C) shows that both the $R$- and $M$-point superlattice peaks have disappeared, and the phase is therefore cubic at this temperature and above (although anisotropic refinement shows the likelihood of localized octahedral tilting – see below). Representative Rietveld plots for these higher temperature phases are provided in Fig. S8.

**LNN-3:**

Our PXRD and PND data at ambient temperature show that LNN-3 occurs exclusively as the polar orthorhombic phase Q (space group $P2_1ma$), in agreement with previous studies[9,11]. This phase persists on raising the temperature as far as 300 °C, with no evidence for any secondary phases. Once again, a check for the possibility of the centrosymmetric alternative *Pnma* phase was made both at 20 °C and 300 °C, but the fits were significantly poorer in each case (at 20 °C: $P2_1ma$: $\chi^2 = 2.44$ for 56 variables; *Pnma*: $\chi^2 = 5.86$ for 43 variables; at 300 °C: $P2_1ma$: $\chi^2 = 2.28$ for 56 variables; *Pnma*: $\chi^2 = 3.49$ for 43 variables).

Lattice parameters throughout the phase Q regime evolve smoothly (see Fig. S9 and S10 for plot of lattice parameters versus $T$ for all compositions).

At 350 °C there is a distinct change in the observed diffraction pattern, noticeable for example in the nature of the peaks near $d \sim 1.95$ Å (Fig. S11)). This indicates a first-order phase transition; which is supported by the observation of a small amount of residual phase Q (e.g. near $d \sim 2.27$ Å). Close inspection of the raw data from the high-resolution detectors (bank 1) shows that both the *M*- and *R*-point superlattice peaks are still present, although there is no clear evidence for the development of any further superlattice. However, inspection of the lower resolution detectors (bank 2) does reveal evidence of additional weak scattering, of a more complex nature, in the region $2.25 < d < 2.50$ Å (Fig. 8). This is apparently not clear in bank 1 simply due to the lower neutron flux within this region (lower signal-noise). Nevertheless, this weak and somewhat broadened additional scattering persists throughout the region $350 \leq T \leq 500$ °C.

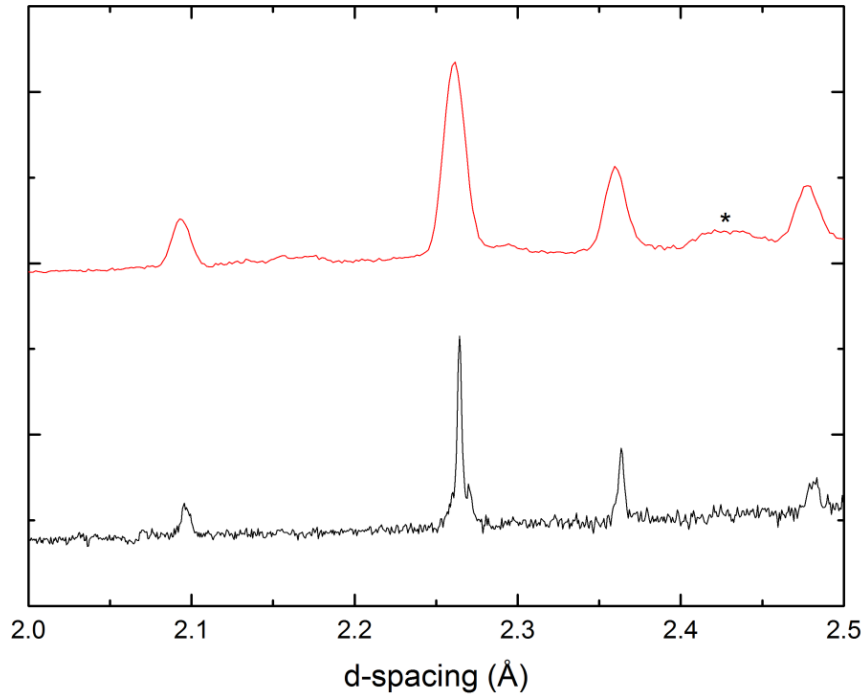

**Figure 8:** Raw data at 350 °C for LNN-3: data measured on bank 1 (black) and bank 2 (red). Note the presence of the broadened superlattice at $2.35 < d < 2.50$ Å, visible only in the bank 2 data.

The most likely explanation for this weak scattering is a further, more complex octahedral tilting pattern, giving rise to additional features along the $T$-line. Initial attempts to fit this scattering using models related to phases R or S of NaNbO$_3$ (i.e. with $4 \times a_p$ or $6 \times a_p$ repeats) were not successful. Nevertheless, it is possible, at least for bank 1, to obtain a good quality of fit with a much simpler model, involving only the standard $M_3^+$ and $R_4^+$ tilts. The simple options are as considered for LNN-12 at 650 °C; $a^0b^+c^-$ (space group *Cmcm*), $a^+a^+c^-$ (space group *P*4$_2$/*nmc*) and $a^-b^+a^-$ (space group *Pnma*). Of these, it is found that the *Cmcm* and *P*4$_2$/*nmc* models produce virtually the same quality of fit ($\chi^2$ values of 3.58 and 3.60, respectively for the 350 °C dataset, Fig. S12). It seems futile to draw further conclusions from this analysis as, despite the very high quality of fit to bank 1, the additional features seen in bank 2 clearly demonstrate that the true structure (and probably the true tilt system) is more complex. Further work is required to establish the exact nature of this phase, for example whether a longer-range or even incommensurate modulation is present. If so, this is certainly distinct from the long-range modulation seen in Phase S´´ for LNN-8, above. Idealized simulations from ISODISTORT suggest possible repeats of $20 \times a_p$ may be required to account for these peaks, if they are indeed due to complex octahedral tilt modes (Fig. S13).

At 550 °C the pattern for bank 2 simplifies again, with the disappearance of the additional weak scattering around the *T*-line. The *M*- and *R*-point peaks still remain, and the optimal model is again found to be that for phase T1 (tilt system $a^0b^+c^-$, space group *Cmcm*). Above this temperature the natural sequence of phase transitions follows: tetragonal phase T2 at 600 °C and cubic phase U at 650 °C to 900 °C. As might be anticipated the corresponding phase transition temperatures for this sequence are somewhat lower than those seen for LNN-12 and LNN-8, due to the higher tolerance factor (i.e. larger average A-site cation size) at this composition. Anisotropic refinement of the oxygen atom in the cubic phase suggests some residual localized octahedral tilting, but with no long-range crystallographic order (Fig. S14).

**Discussion**

Previous studies of the high temperature phase behavior of the $Li_xNa_{1-x}NbO_3$ system have focused mainly on dielectric and pyroelectric measurements, generally for low Li contents (x < 0.20)[21-23]. Crystallographic studies have been limited to those on the end members $NaNbO_3$[6,7,24] and $LiNbO_3$[8], and the two recent studies of LNN-12[14] and LNN-20[12] compositions. Mishra's study of LNN-12[14] suggests the following sequence of phases (based on the notation used above): (Na-R3c + Q) → Q → (Q + T1) → T1 → T2 → U, with the corresponding temperatures being approximately 400, 500, 650, 920 and 1000 K, respectively. Although also based on both X-ray and neutron power diffraction, this study did not report the presence of Phase S´; the significantly enhanced resolution of the powder neutron diffraction data used in the present study is the key to identification of this novel phase.

$NaNbO_3$ itself has perhaps the most complex temperature-dependent phase diagram of any $ABX_3$ perovskite, whereas $LiNbO_3$ retains adopts only one phase below $T_C$ (~1200 °C). Prior to the present work, therefore it was difficult to predict what phases might be present in the high temperature regime of the LNN system. Based on the analysis above, the phases we observe in this study, combined with that in ref. 12 are summarized in Fig. 9. It is interesting to note that the relatively well-defined $4 \times a_p$ and $6 \times a_p$ superlattices seen in phases S and R of $NaNbO_3$[10] are already lost at 3% Li-doping (LNN-3), giving way to a more diffuse long-range superlattice, which apparently changes again at LNN-8, only to re-organise to a well-defined (but different) $4 \times a_p$ superlattice at LNN-12. The long-range tilt phases disappear at LNN-20 which, although also displaying a unique phase sequence, does lie within the established Glazer family of tilt systems, which are based on a $2 \times 2 \times 2$ array of $BO_6$ octahedra.

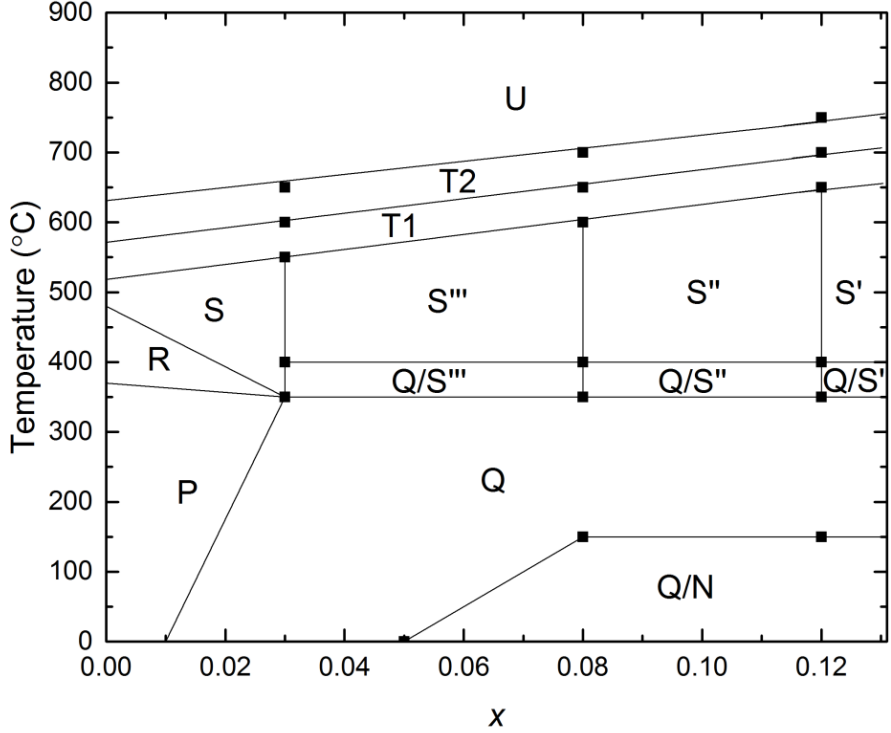

**Figure 9:** Schematic phase diagram for LNN-X, incorporating previously reported phases from refs **6,7** The black squares indicate phase boundaries identified in the present study.

Previous studies have noted the tendency of 'softening' of perovskite tilt modes along the *T*-line; indeed several examples of diffuse streaking along ***k*** = (½, ½, γ) in electron diffraction patterns have been observed, for example in $Ca_{0.37}Sr_{0.63}TiO_3$[25] and in $NaNbO_3$ itself[26]. Discrete long-range superstructures have also been proposed based on powder diffraction studies, for example a $\sqrt{2}\ a_p \times 14\sqrt{2}\ a_p \times 2\ a_p$ superlattice has been proposed in the $NaNbO_3$-$CaTiO_3$ system[27] and a $10\ a_p \times 10\ a_p \times 2\ a_p$ superlattice in $KLaMnWO_6$[28]. In the latter case the superlattice has been suggested to occur due to cooperative octahedral tilting and A-site cation compositional modulation, leading to so-called 'octahedral twinning,' in which two $5 \times a_p$ repeat of out-of-phase tilts are 'twinned' through a single in-phase tilt. It is difficult to imagine that such a cation –induced modulation could occur in the present case, as there is no obvious driver towards Na/Li ordering, as there is with the charge-ordered $K^+/La^{3+}$ case (although we suggested some tentative evidence for this in the high-temperature phases of LNN-20[12]). Moreover, it is also difficult to speculate on the exact nature of the octahedral tilting in the long-range superstructures postulated in LNN-3 and LNN-8. They needn't be based on simple *irreps* of the type shown in Figs. S2 and S7, but could involve compound tilts incorporating simultaneous $T_4/M_3^+/R_4^+$ tilts along one axis, or even along more than one axis.

**Conclusions**

We present here the first thorough and detailed study of the high temperature phase behavior of $Li_xNa_{1-x}NbO_3$ within the Na-rich composition region (x < 0.20). Each of the three distinct compositions studied, *viz.* LNN-3, LNN-8 and LNN-12, displays evidence for unique long-range structural modulations based on complex octahedral tilt schemes. We find clear evidence for one distinct and novel complex octahedral tilt phase for a composition $Li_{0.12}Na_{0.88}NbO_3$ (LNN-12) in the temperature regime 400 to 600 °C. This phase (S´) is similar to Phase S of $NaNbO_3$, in the sense that both exhibit $2\,a_p \times 2\,a_p \times 4\,a_p$ supercells of the aristotype perovskite unit cell, but the octahedral tilt patterns displayed by the two phases are different. We also show tentative evidence for yet more complex tilt phases in compositions LNN-8 and LNN-3 at elevated temperatures. This study highlights the importance (but also the limitations) of *high-resolution* powder neutron diffraction in the study of such complex phases Further experiments (high-temperature electron diffraction or single crystal X-ray or neutron diffraction) and perhaps theoretical work will be necessary in order to verify whether our suggestion of an unprecedented $30 \times a_p$ superlattice is borne out. In any case, we have shown that the remarkable structural complexity shown by $NaNbO_3$ itself is carried forward and extended further in its solid solution with $LiNbO_3$. The underlying causes of the apparent structural frustration in these systems remains to be understood.

**Supporting Information**

See Supplemental Material at xxx for further crystallographic details, including refined models in CIF format

**Acknowledgements**

We thank STFC for provision of neutron diffraction facilities at ISIS and Dr Aziz Daoud-Aladine for experimental assistance. CALD was supported by an EPSRC DTA studentship (EP/L50579/1).

Research data supporting this publication can be accessed at xxxxx.

**TOC Graphic:**

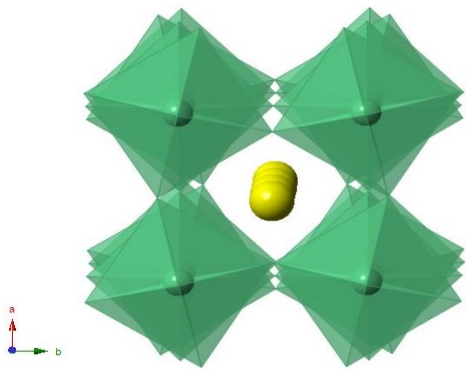 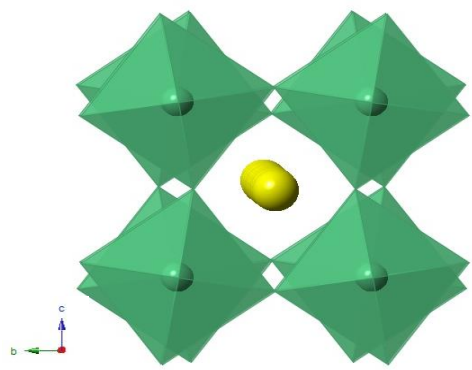